\documentclass[runningheads,a4paper]{llncs}

\usepackage{amssymb}
\usepackage{graphicx}

\usepackage{url}
\urldef{\mailsa}\path|{alfred.hofmann, ursula.barth, ingrid.haas, frank.holzwarth,|
\urldef{\mailsb}\path|anna.kramer, leonie.kunz, christine.reiss, nicole.sator,|
\urldef{\mailsc}\path|erika.siebert-cole, peter.strasser, lncs}@springer.com|    
\newcommand{\keywords}[1]{\par\addvspace\baselineskip
\noindent\keywordname\enspace\ignorespaces#1}

\usepackage{graphics} 
\usepackage{epsfig} 

\begin{document}

\mainmatter  

\title{Determining the Number of Holes of  a 2D Digital Component is Easy}

\titlerunning{Digital Curvatures Applied to a 3D Object}

\author{Li Chen\inst{1} }

\authorrunning{L. Chen}

\institute{University of the District of Columbia\\ \email{lchen@udc.edu}  
}

\index{Chen, Li}

\maketitle

\begin{abstract}
The number of holes in a connected component in 2D images is a basic invarient. 
In this note, a simple formula was proven using our previous results in digital 
topology~\cite{Che04,CR}. The new is: $h =1+ (|C_4|-|C_2|)/4$ , where h is the number of holes, and
$C_i$ indicate the set of corner points having $i$ direct adjacent points in the component.

\end{abstract}


\section{Introduction}

An image segmentation method can extract a connected component.  
A connected component $S$ in a 2D digital image is often used to represent
a real object. The identification of the object can be first done by determining
how many holes in the component. For example, letter ``A'' has one hole
and ``B'' has two holes. In other words, if $S$ has $h$ holes, then the complement of $S$ 
has $h+1$ connected components (if S does not reach the boundary of the image).

In this note, we provide two proofs for the following formula: 
\[ h =1+ (|C_4|-|C_2|)/4   \] 
\noindent  where h is the number of holes, and
$C_i$ indicate the set of corner points having $i$ direct adjacent points in the component.

\section{Some Concepts and Definitions of Digital Space }

A digital space  is a discrete space in which each
point can be defined as an integer vector.  

Two-dimensional digital space $\Sigma_{2}$ first. A point P (x, y) in $\Sigma_{2}$ has two
horizontal $(x, y\pm 1)$ and two vertical neighbors $(x\pm 1, y)$. These four
neighbors are called directly adjacent points of $p$ . $p$ has also four
diagonal neighbors: $(x\pm 1, y\pm 1)$. These eight (horizontal, vertical and
diagonal) neighbors are called general (or indirect)
adjacent
points of $p$ .

        Let $\Sigma_{m}$ be m-dimensional digital space. Two points
$p=(x_1, x_2,...,x_m)$ and $q=(y_1,y_2,...,y_m)$ in $\Sigma_{m}$ are
directly adjacent points, or we say that $p$ and $q$ are direct
neighbor if

\centerline {$ d_{D}(p,q)=\sum_{i=1}^m |x_i-y_i|=1 .$}

\noindent $p$ and $q$ are indirectly adjacent points if

\centerline {$ d_{I}(p,q)=\max_{1\le i\le m} |x_i-y_i|=1. $}

\noindent Note: ``Indirectly adjacent points'' include all
directly adjacent points here. It may be the reason that we should
change the word of ``indirectly'' to ``generally.''

        In a three-dimensional space $\Sigma_{3}$, a point has six directly
adjacent points and 26 indirectly adjacent points. Therefore, two directly
adjacent points in $\Sigma_{3}$ are also called 6-connected, while two
indirectly adjacent points are also called
26-connected.
In this note, we mainly consider the direct adjacency. If we omit the word ``direct,'' ``adjacency'' means the direct adjacency.

        A point in $\Sigma_{m}$ is called a point-cell or
0-cell.
A pair of points
$\{p,q\}$ in $\Sigma_{m}$ is called a line-cell\index{Line-cell} or
1-cell\index{1-cell}, if $p$ and $q$ are adjacent
points. A surface-cell\index{surface-cell} is a set of 4 points which form a unit square
parallel to coordinate planes. A 3-dimensional-cell (or 3-cell)\index{3-cell}
is a unit cube which includes 8 points. By the same reasoning,
we may define a $k$-cell.  Fig. 2.1(a)(b)(c)(d) show a point-cell, line-cell, a
surface-cell and a 3-cell, respectively.

        Now let us consider to the concepts of adjacency and connectedness of
(unit) cells. Two points $p$ and $q$ (point-cells, or 0-cells) are
connected if there exists a simple path $p_0,p_1,...,p_n$, where $p_0=p$ and
$p_n=q$, and $p_i$ and $p_{i+1}$ are adjacent for $i=1,...,n-1 $.

       Two cells are point-adjacent\index{Point adjacent} if they share a point. For example,
line-cells $C1$ and $C2$ are point-adjacent in Fig. 2.1 (e), and
surface-cells $s1$ and $s2$ are point-adjacent in Fig. 2.1(f).
Two surface-cells are line-adjacent\index{Line adjacent} if they share a line-cell. For
example, surface-cells $s1$ and $s3$ in Fig. 2.1(g) are line-adjacent.

       Two line-cells are point-connected\index{Point connected} if they are two end elements
of a line-cells path in which each pair of adjacent line-cells
is point-adjacent. For example, line-cells $C1$ and $C3$ in Figure
2.1 (e) are point-connected. Two surface-cells are
 line-connected\index{Line connected} if they are two end elements of a surface-cells path
 in which each pair adjecent surface-cells
are point-adjacent.
For example, $s1$ and $s2$ in Fig. 2.1(f) are line-connected.

Two $k$-cells are $k'$-dimensional adjacent ($k'$-adjacent),
$k>k'\ge 0$, if they share a $k'$-dimensional cell.
A (simple) $k$-cells path
with $k'$-adjacency is a sequence of  $k$-cells $v_0,v_1,...,v_n$,
 where $v_i$ and $v_{i+1}$
 are $k'$-adjacent and  $v_0,v_1,...,v_n$
are different elements. Two $k$-cells are called $k'$-dimensional
connected if
they are two end elements of a (simple) $k$-cells path
with $k'$-adjacency.

Assume that $S$ is a subset of $\Sigma_{m}$.
Let $\Gamma^{(0)}(S)$ be the set of all points in $S$,
and $\Gamma^{(1)}(S)$ be the line-cells set in $S$,...,
$\Gamma^{(k)}(S)$\index{$\Gamma^{(k)}(S)$} be the set of $k$-cells of $S$.
We say two elements $p$ and $q$ in  $\Gamma^{(k)}(S)$ are $k'$-adjacent
if $p\cap q\in \Gamma^{(k')}(S)$, $k' < k$.

	Let $p\in \Sigma_{3}$, a line-neighborhood of $p$ is a set containing $p$
and its two adjacent points. A surface-neighborhood of $p$ is
 a (sub-)surface where
$p$ is a inner point of the (sub-)surface.

$\Sigma_{m}$ represents a special graph $\Sigma_{m}=(V,E)$.
$V$ contains all integer
grid points in the $m$ dimensional Euclidean space \cite{MR,Che04}.
The edge set $E$ of $\Sigma_{m}$
is defined as $E = \{(a,b) | a, b \in V \& d(a,b)=1\}$ , where $d(a,b)$
is the distance between $a$ and $b$. In fact, $E$ contains all
pairs of adjacent points.
Because $a$ is an $m$-dimensional vector, $(a,b)\in E$
means that only
one component, the $i$-th component, is different in $a$ and $b$,
$|x_i - y_i|=1$, and the
rest of the components are the same where
 $a=(x_1,...,x_m)$ and  $b =(y_1,...,y_m)$. This is known as
the direct adjacency. One can define indirect adjacency as
$\max_{i} |x_i - y_i|=1 $.
 $\Sigma_{m}$  is usually called an $m$-dimensional digital space.
The basic discrete geometric element $n$-cells can be defined in such a
space, such as 0-cells (point-cells), 1-cells (line-cells), and 2-cells
(surface-cells).


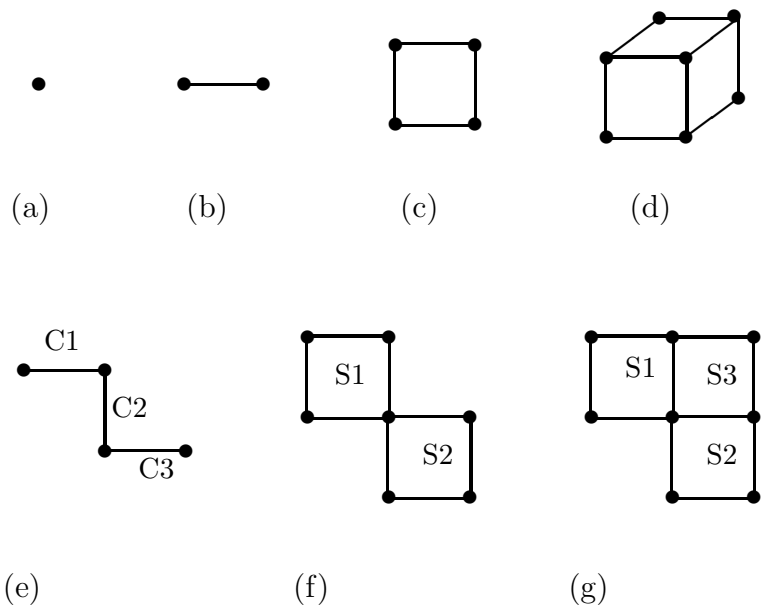
\begin{figure}[hbt]

\begin{center}
\setlength{\unitlength}{2500sp}%
\begingroup\makeatletter\ifx\SetFigFont\undefined%
\gdef\SetFigFont#1#2#3#4#5{%
  \reset@font\fontsize{#1}{#2pt}%
  \fontfamily{#3}\fontseries{#4}\fontshape{#5}%
  \selectfont}%
\fi\endgroup%
\begin{picture}(7448,5872)(2176,-6565)
\thicklines
\put(2531,-1446){\circle*{132}}
\put(3956,-1446){\circle*{132}}
\put(4732,-1446){\circle*{132}}
\put(8101,-1185){\circle*{132}}
\put(8101,-1970){\circle*{132}}
\put(8878,-1970){\circle*{132}}
\put(8878,-1185){\circle*{132}}
\put(9396,-1577){\circle*{132}}
\put(8620,-792){\circle*{132}}
\put(6028,-1054){\circle*{132}}
\put(6806,-1054){\circle*{132}}
\put(6806,-1839){\circle*{132}}
\put(6028,-1839){\circle*{132}}
\put(9351,-773){\circle*{132}}
\put(8101,-1970){\framebox(777,785){}}
\put(6028,-1839){\framebox(778,785){}}
\put(8878,-1185){\makebox(6.6667,10.0000){\SetFigFont{10}{12}{\rmdefault}{\mddefault}{\updefault}.}}
\put(8878,-1970){\makebox(6.6667,10.0000){\SetFigFont{10}{12}{\rmdefault}{\mddefault}{\updefault}.}}
\put(8101,-1185){\line( 4, 3){520.800}}
\put(8620,-792){\line( 1, 0){776}}
\put(9396,-792){\line( 0,-1){785}}
\put(9396,-1577){\line(-4,-3){520.160}}
\put(9396,-792){\line(-4,-3){520.160}}
\put(3956,-1446){\line( 1, 0){776}}
\put(7953,-3944){\circle*{132}}
\put(8748,-3944){\circle*{132}}
\put(8748,-4737){\circle*{132}}
\put(7953,-4737){\circle*{132}}
\put(9543,-4737){\circle*{132}}
\put(8748,-5532){\circle*{132}}
\put(9543,-5532){\circle*{132}}
\put(9543,-3944){\circle*{132}}
\put(7953,-4737){\framebox(795,793){}}
\put(8748,-5532){\framebox(795,795){}}
\put(8748,-3944){\line( 1, 0){729}}
\put(9477,-3944){\line( 1, 0){ 66}}
\put(9543,-3944){\line( 0,-1){793}}
\put(5169,-3944){\circle*{132}}
\put(5963,-3944){\circle*{132}}
\put(5963,-4737){\circle*{132}}
\put(5169,-4737){\circle*{132}}
\put(6759,-4737){\circle*{132}}
\put(5963,-5532){\circle*{132}}
\put(6759,-5532){\circle*{132}}
\put(5169,-4737){\framebox(794,793){}}
\put(5963,-5532){\framebox(796,795){}}
\put(2384,-4274){\circle*{132}}
\put(3179,-4274){\circle*{132}}
\put(3179,-5069){\circle*{132}}
\put(3974,-5069){\circle*{132}}
\put(2384,-4274){\line( 1, 0){795}}
\put(3179,-4274){\line( 0,-1){795}}
\put(3179,-5069){\line( 1, 0){795}}
\put(2583,-4075){\makebox(0,0)[lb]{\smash{\SetFigFont{11}{13.2}{\familydefault}{\mddefault}{\updefault}C1}}}
\put(3245,-4737){\makebox(0,0)[lb]{\smash{\SetFigFont{11}{13.2}{\familydefault}{\mddefault}{\updefault}C2}}}
\put(3511,-5334){\makebox(0,0)[lb]{\smash{\SetFigFont{11}{13.2}{\familydefault}{\mddefault}{\updefault}C3}}}
\put(5433,-4407){\makebox(0,0)[lb]{\smash{\SetFigFont{11}{13.2}{\familydefault}{\mddefault}{\updefault}S1}}}
\put(6295,-5202){\makebox(0,0)[lb]{\smash{\SetFigFont{11}{13.2}{\familydefault}{\mddefault}{\updefault}S2}}}
\put(9079,-5202){\makebox(0,0)[lb]{\smash{\SetFigFont{11}{13.2}{\familydefault}{\mddefault}{\updefault}S2}}}
\put(9079,-4407){\makebox(0,0)[lb]{\smash{\SetFigFont{11}{13.2}{\familydefault}{\mddefault}{\updefault}S3}}}
\put(8283,-4341){\makebox(0,0)[lb]{\smash{\SetFigFont{11}{13.2}{\familydefault}{\mddefault}{\updefault}S1}}}
\put(2251,-2761){\makebox(0,0)[lb]{\smash{\SetFigFont{12}{14.4}{\familydefault}{\mddefault}{\updefault}(a)  }}}
\put(2176,-6511){\makebox(0,0)[lb]{\smash{\SetFigFont{12}{14.4}{\familydefault}{\mddefault}{\updefault}(e)  }}}
\put(3976,-2761){\makebox(0,0)[lb]{\smash{\SetFigFont{12}{14.4}{\familydefault}{\mddefault}{\updefault}(b)  }}}
\put(6076,-2761){\makebox(0,0)[lb]{\smash{\SetFigFont{12}{14.4}{\familydefault}{\mddefault}{\updefault}(c)  }}}
\put(8326,-2761){\makebox(0,0)[lb]{\smash{\SetFigFont{12}{14.4}{\familydefault}{\mddefault}{\updefault}(d) }}}
\put(5026,-6511){\makebox(0,0)[lb]{\smash{\SetFigFont{12}{14.4}{\familydefault}{\mddefault}{\updefault}(f)  }}}
\put(7726,-6511){\makebox(0,0)[lb]{\smash{\SetFigFont{12}{14.4}{\familydefault}{\mddefault}{\updefault}(g)   }}}
\end{picture}
\end{center}
\caption{Examples of basic unit cells and their connections :
    (a) 0-cells, (b) 1-cells, (c) 2-cells, (d) 3-cells, (e) point-connected
    1-cells, (f) point-connected 2-cells, and (f) line-connected 2-cells.}
\end{figure}

\section{Two Previous Related Results}

We have proved some related theorem using Euler Characteristics and Gauss-Bonett Theorem.
The first is about simple closed digital curves. 

$C$ is a simple closed curve where
each element in $C$ is a point in  $\Sigma_{2}$.
In addition, $C$ does not contain the following
cases:

$\begin{array}{ll}
1 & 0 \\
0 & 1
\end{array} $

\noindent and

$\begin{array}{ll}
0 & 1 \\
1 & 0
\end{array}$

\noindent These two cases are called  the pathelogical 
cases.

We use $IN_{C}$ to represent the internal part of $C$. 
Since direct adjacency has the Jordan separation property,
$\Sigma_2-C$ will be disconnected. 

We also call a point $p$ on $C$ a $CP_{i}$ point if $p$ has
$i$ adjacent points in $IN_{C}\cup C$. In fact, $|CP_{1}|=0$ and
$|CP_{i}|=0$ if $i>4$ in $C$.

$CP_{2}$ contains outward corner points, $CP_{3}$ contains straight-line points, and
$CP_{4}$ contains inward corner points. For example, the following center point is
a outward corner point:

$\begin{array}{lll}
0 & 0 & 0 \\
0 & 1 & 1 \\
0 & 1 & x 
\end{array} $

\noindent But in next array, the center point is an inward corner point:  

$\begin{array}{lll}
0 & 1 & x \\
1 & 1 & x \\
x & x & x 
\end{array} $

In \cite{Che04}, we showed for $C$,
 
\begin{lemma} 
\begin{equation}
      CP_2 = CP_4 + 4.
\end{equation}
\end{lemma}

For a 3D image,   Since 
cubical space with direct adjacency, or (6,26)-connectivity space, has the simplest 
topology in 3D digital spaces, we will use it as the 3D image domain. 
It is also believed to be sufficient for the topological 
property extraction of digital objects in 3D.  
In this space, two points are said to be adjacent in 
(6,26)-connectivity space if the 
Euclidean distance between these two points is 1.
 
Let $M$ be a closed (orientable) digital surface in $\Sigma_3$. 
in direct adjacency. 
We know that there are exactly 6-types of digital surface
points~\cite{Che04}\cite{CR}. 

\begin{figure}[hbt]
   {
	\begin{center}
		\setlength{\unitlength}{0.008in}%
 
\begin{picture}(375,560)(60,260)

\thicklines
\put(275,680){\line( 1, 1){ 20}}
\put(295,700){\line( 1, 0){ 60}}
\put(355,700){\line( 0,-1){ 60}}
\put(355,640){\line(-1,-1){ 20}}
\put(355,700){\line(-1,-1){ 20}}
\put(275,680){\line( 1, 0){ 60}}
\put(335,680){\line( 0,-1){ 60}}
\put(295,700){\line( 1, 1){ 20}}
\put(315,720){\line( 1, 0){ 60}}
\put(375,720){\line( 0,-1){ 60}}
\put(375,660){\line(-1,-1){ 20}}
\put(375,720){\line(-1,-1){ 20}}
\put( 80,380){\line(-1,-1){ 20}}
\put( 60,360){\line( 1, 0){ 60}}
\put(120,360){\line( 1, 1){ 20}}
\put(140,380){\line(-1, 0){ 60}}
\put(120,360){\line( 1, 1){ 20}}
\put( 80,380){\line( 1, 1){ 20}}
\put(260,360){\line( 1, 1){ 20}}
\put(280,380){\line( 1, 0){ 60}}
\put(340,380){\line( 0,-1){ 60}}
\put(340,320){\line(-1,-1){ 20}}
\put(340,380){\line(-1,-1){ 20}}
\put(260,360){\line( 1, 0){ 60}}
\put(320,360){\line( 0,-1){ 60}}
\put(140,380){\line( 1, 0){ 60}}
\put(200,380){\line( 0,-1){ 60}}
\put(200,320){\line(-1, 0){ 60}}
\put(280,380){\line( 0, 1){ 60}}
\put(280,440){\line( 1, 0){ 60}}
\put(340,440){\line( 0,-1){ 60}}
\put(140,440){\line( 1, 0){ 60}}
\put(200,440){\line( 0,-1){ 60}}
\put(340,440){\line( 1, 1){ 20}}
\put(360,460){\line( 0,-1){ 60}}
\put(120,360){\line( 0,-1){ 60}}
\put(120,300){\line( 1, 1){ 20}}
\put(140,320){\line( 0, 1){ 60}}
\put(140,380){\line( 1, 1){ 20}}
\put(160,400){\line(-1, 0){ 60}}
\put(140,380){\line( 0, 1){ 60}}
\put(140,440){\line( 1, 1){ 20}}
\put(160,460){\line( 0,-1){ 60}}
\put(340,380){\line( 1, 0){ 60}}
\put(400,380){\line( 0,-1){ 60}}
\put(400,320){\line(-1, 0){ 60}}
\put(340,380){\line( 1, 1){ 20}}
\put(360,400){\line( 1, 0){ 60}}
\put(420,400){\line(-1,-1){ 20}}
\put( 60,560){\line( 1, 1){ 20}}
\put( 80,580){\line( 1, 0){ 60}}
\put(140,580){\line( 0,-1){ 60}}
\put(140,520){\line(-1,-1){ 20}}
\put(140,580){\line(-1,-1){ 20}}
\put( 60,560){\line( 1, 0){ 60}}
\put(120,560){\line( 0,-1){ 60}}
\put(140,580){\line( 1, 0){ 60}}
\put(200,580){\line( 0,-1){ 60}}
\put(200,520){\line(-1, 0){ 60}}
\put(140,580){\line( 1, 1){ 20}}
\put(160,600){\line( 1, 0){ 60}}
\put(220,600){\line(-1,-1){ 20}}
\put( 80,580){\line( 1, 1){ 20}}
\put(100,600){\line( 1, 0){ 60}}
\put( 80,680){\line(-1,-1){ 20}}
\put( 60,660){\line( 1, 0){ 60}}
\put(120,660){\line( 1, 1){ 20}}
\put(140,680){\line(-1, 0){ 60}}
\put(120,660){\line( 1, 0){ 60}}
\put(180,660){\line( 1, 1){ 20}}
\put(200,680){\line(-1, 0){ 60}}
\put( 80,680){\line( 1, 1){ 20}}
\put(100,700){\line( 1, 0){120}}
\put(220,700){\line(-1,-1){ 20}}
\put(160,700){\line(-1,-1){ 20}}
\put( 60,800){\line( 1, 0){ 60}}
\put(120,800){\line( 0,-1){ 60}}
\put( 60,800){\line( 1, 1){ 20}}
\put( 80,820){\line( 1, 0){ 60}}
\put(140,820){\line( 0,-1){ 60}}
\put(140,760){\line(-1,-1){ 20}}
\put(140,820){\line(-1,-1){ 20}}
\put( 80,820){\line( 0,-1){ 60}}
\put( 80,760){\line( 1, 0){ 60}}
\put( 95,725){\makebox(0,0)[lb]{\raisebox{0pt}[0pt][0pt]{ (1)}}}
\put(110,625){\makebox(0,0)[lb]{\raisebox{0pt}[0pt][0pt]{ (2)}}}
\put(305,625){\makebox(0,0)[lb]{\raisebox{0pt}[0pt][0pt]{ (3)}}}
\put(125,485){\makebox(0,0)[lb]{\raisebox{0pt}[0pt][0pt]{ (4)}}}
\put(125,285){\makebox(0,0)[lb]{\raisebox{0pt}[0pt][0pt]{ (5)}}}
\put(325,285){\makebox(0,0)[lb]{\raisebox{0pt}[0pt][0pt]{ (6)}}}
\end{picture}
	\end{center}
	}

\caption{Six types of digital surfaces points in 3D.}
\end{figure}
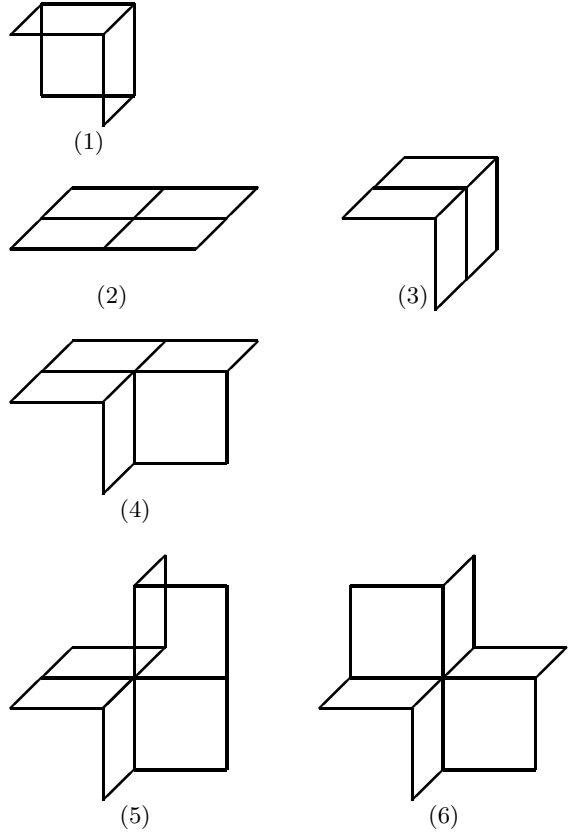

Assume that $M_i$ ($M_3$, $M_4$, $M_5$, $M_6$) is the set of 
digital points with $i$ 
neighbors. We have the following result for a simply connected 
$M$ ~\cite{Che04}:
\begin{equation}
          |M_3| =8 + |M_5| + 2 |M_6| .      
\end{equation}
 
\noindent We also have a genus formula based on the Gauss-Bonnet Theorem \cite{CR}

\begin{equation}
           g = 1+ (|M_5|+2 \cdot |M_6| -|M_3|)/8. 
\end{equation} 

\section{The Simple Formula for the Number of Holes in $S$} 

In this section, we first use the 3D formula to get the 
theorem for holes. 
 
Let $S\subset \Sigma_2$ be a connected component and its boundary do not have the pathelogical 
cases.  (We actually can detect those cases in linear time.)

We can embed $S$ into $\Sigma_3$ to make a double $S$ in $\Sigma_3$. 
At $z=1$ plane, we have a $S$, denoted $S_1$, and we also have the exact same
$S$ at $z=2$ plane, denoted $S_2$. 

Without loss generality, $S_1\cup S_2$ is a solid object. (We here omit some
technical details for the strict definition of digital surfaces.) It's boundary
is closed digital surfaces with genus $g=h$. We know   
$g = 1+ (|M_5|+2 \cdot |M_6| -|M_3|)/8$. 

There will be no points in $M_6$. We have

\begin{theorem}
Let $S \subset \Sigma_2$ be a connected component and its boundary $B$ is a
collection of simple closed curves  without pathelogical cases.  
Then, the number of holes in $S$ is
\begin{equation}
   h= 1+(C_4-C_2)/4
\end{equation}
\noindent  $C_4, C_2 \subset B$.
\end{theorem}

{\it Proof: } 
For each point $x$ in $C_2$ in $C\subset S$ ($C$ is the boundary of $S$), 
we will get
two points in $M_3$ in $S_1\cup S_2$. In the same way, if a point $y$ is
inward in $C_4\in C$, we will get
two points in $M_5$ in $S_1\cup S_2$. There is no point in $M_6$, i,e.,
$|M_6|=0$. 
So $2|C_2|=|M_3|$, and $2|C_4|=|M_5|$. We have

$h=g = 1+ (|M_5|+2 \cdot |M_6| -|M_3|)/8= 1+ (2|C_4|-2|C_2|)/8 $

\noindent Thus,

$h= 1+ (|C_4|-|C_2|)/4 $.

We can also prove this theorem using the curve theorem:
$CP_2=CP_4+4$ for a simple closed curve. 

{\it The Second Proof: }
  
This can also be proved by the lemma in above section. 
$CP_2=CP_4+4$;  

A 2D connected component $S$ with $h$ holes that contains 
$h+1$ simple closed curves in the boundary of $S$
Those curves do not cross each other. 

The $h$ curves correspoding to $h$ holes will be considered 
oppositely in terms of inward-outward. 

including one counts at inward and h is reversed
outward with inward. It will get there. 

Let $CP^{(0)}$ the outside curve of $S$ and  
$CP^{(i)}, i=1,\cdots,h$, is the curve for the $i$-th hole.  

Inward points to $S$ is the outward points to $C^{(i)}, i=1,\cdots,h$.
And vise versa.

$CP_2^{(0)}=CP_4^{(0)}+4$

$CP_2^{(i)}=CP_4^{(i)}+4$

The total outward points in the boundary of $S$ is

 $CP_{2}=CP_2^{(0)} + \sum_{i=1}^{h} CP_4^{(i)}$.

\noindent The inward points in the boundary of $S$ is
  
 $CP_{4}=CP_4^{(0)} + \sum_{i=1}^{h} CP_2^{(i)}$

\noindent Thus,
 $CP_{4}-CP_{2}=CP_4^{(0)} + \sum_{i=1}^{h} CP_2^{(i)} 
                -CP_2^{(0)} - \sum_{i=1}^{h} CP_4^{(i)}$

we have 
 $CP_{4}-CP_{2}=-4  + \sum_{i=1}^{h} 4  = -4 +4h$
 
Therefore, 

 $h = 1 + (CP_{4}-CP_{2})/4$.

Therefore this formula is so simple to get the holes (genus) for a 2D object
without any little sophisted algorithm, just count if the point is
a corner point, inward or outward. 

We could not get the simular simple formula in triangulated 
representation of the 2D object. This is the beauty of digital 
geometry and topology!

To test if this formula is correct, we can select the following examples

\begin{eqnarray}
    \left (\begin{array}{llllllll}

0 & 0 & 0 & 0 & 0 & 0 & 0 & 0\\
0 & 0 & 1 & 1 & 1 & 1 & 0 & 0   \\
0 & 1 & 1 & 1 & 1 & 1 & 0 & 0 \\
0 & 1 & 1 & 1 & 0 & 0 & 0 & 0\\
0 & 0 & 1 & 1 & 0 & 0 & 0 & 0   \\
0 & 0 & 1 & 1 & 1 & 0 & 0 & 0\\
0 & 0 & 1 & 1 & 1 & 0 & 0 & 0   \\
0 & 0 & 0 & 0 & 0 & 0 & 0 & 0
     
              \end{array}
              \right )
\end{eqnarray}

In order to see clearly, we use ``2'' to represent points in $CP_2$ and
use ``4'' to represent points in $CP_4$.

\begin{eqnarray}
    \left (\begin{array}{llllllll}

0 & 0 & 0 & 0 & 0 & 0 & 0 & 0\\
0 & 0 & 2 & 1 & 1 & 2 & 0 & 0   \\
0 & 2 & 4 & 4 & 1 & 2 & 0 & 0 \\
0 & 2 & 4 & 1 & 0 & 0 & 0 & 0\\
0 & 0 & 1 & 1 & 0 & 0 & 0 & 0   \\
0 & 0 & 1 & 4 & 2 & 0 & 0 & 0\\
0 & 0 & 2 & 1 & 2 & 0 & 0 & 0   \\
0 & 0 & 0 & 0 & 0 & 0 & 0 & 0
     
              \end{array}
              \right )
\end{eqnarray}

In this example $|CP_2| = 8$ and $|CP_4| = 4$.  
 $h = 1 + (CP_{4}-CP_{2})/4 = 1 + (4-8)/4 =0$.

Another example is the following

\begin{eqnarray}
    \left (\begin{array}{llllllll}

0 & 0 & 0 & 0 & 0 & 0 & 0 & 0\\
0 & 0 & 1 & 1 & 1 & 1 & 1 & 1   \\
0 & 1 & 1 & 1 & 1 & 1 & 1 & 1 \\
0 & 1 & 1 & 1 & 0 & 0 & 1 & 1\\
0 & 1 & 1 & 1 & 0 & 0 & 1 & 1   \\
0 & 0 & 1 & 1 & 1 & 1 & 1 & 1\\
0 & 0 & 1 & 1 & 1 & 1 & 1 & 1   \\
0 & 0 & 0 & 0 & 0 & 0 & 0 & 0
     
              \end{array}
              \right )
\end{eqnarray}

In the second example $|CP_2| = 6$ and $|CP_4| = 6$.  
 $h = 1 + (CP_{4}-CP_{2})/4 = 1 + (6-6)/4 =1$.
 
\noindent When add a hole, we will add 4 more $CP_4$ points. That is the
reason why this formula is correct.

\section{Conclusion} 

In this paper, we have used digital topology to 
get a simple formula for calculating the number of holes
in a connected component in 2D digital space.
The formula is so simple and can be easily implemented.
The author does not know if this formula was known
or obtained already by other researchers.

\end{document}